\documentclass[reprint,
superscriptaddress,
 amsmath,amssymb,
 aps,
]{revtex4-2}
\usepackage{graphicx}% Include figure files
\usepackage{float}
\usepackage{dcolumn}% Align table columns on decimal point
\usepackage{bm}% bold math
\usepackage{hyperref}
\usepackage{amsmath}
\usepackage{xcolor}

\hypersetup{hypertex=true,
            colorlinks=true,
            linkcolor=blue,
            anchorcolor=blue,
            citecolor=blue,
            urlcolor=blue,        
           }

\begin{document}

\preprint{APS/123-QED}

\title{The coexistence of possible magnetic and chiral rotation in $^{129}\mathrm{Cs}$ and $^{131}\mathrm{La}$: a microscopic investigation }% Force line breaks with \\
\author{Jia-nuo Zang}
\affiliation{College of Physics, Jilin University, Changchun 130012, China}
\author{Duo Chen}
\affiliation{School of Physics, Harbin Institute of Technology, Harbin 150001, China.}%Lines break 
\author{Rui Guo}
\affiliation{Institute of Modern Physics, Chinese Academy of Sciences, Lanzhou 730000, China}
\author{Jian Li}
\email{E-mail:jianli@jlu.edu.cn}
\author{Dong Yang}
\email{E-mail:dyang@jlu.edu.cn }
\affiliation{College of Physics, Jilin University, Changchun 130012, China}
\author{Yue Shi}
\affiliation{School of Physics, Harbin Institute of Technology, Harbin 150001, China.}

%\affiliation{
%\affiliation{
%\affiliation{%


\date{\today}% It is always \today, today,

\begin{abstract}
A microscopic investigation of the rotational properties in $^{129}\mathrm{Cs}$ and $^{131}\mathrm{La}$ was carried out using the three-dimensional tilted axis cranking covariant density functional theory (3DTAC-CDFT). The calculations reveal the coexistence of magnetic and chiral rotation built on identical qusiparticle configurations $\pi {h}_{11/2}^{1}\otimes \nu {h}_{11/2}^{-2}$ in $^{129}\mathrm{Cs}$ and $^{131}\mathrm{La}$, establishing a new type of shape coexistence. The calculations predict the deformation parameters for magnetic rotation in $^{129}\mathrm{Cs}$ ($\beta\approx0.23$, $\gamma \approx41^\circ$) and $^{131}\mathrm{La}$ ($\beta \approx0.25,$  $\gamma \approx42^\circ$), along with those for possible chiral rotation in $^{129}\mathrm{Cs}$ ($\beta \approx0.20,$  $\gamma \approx29^\circ$) and in $^{131}\mathrm{La}$ ($\beta \approx0.20,$  $\gamma \approx27^\circ$). The energy spectra, the relation between the spin and the rotational frequency, and the reduced $M1$ and $E2$ transition probabilities are obtained with the various configurations. The experimental characteristics of band B8 in $^{129}\mathrm{Cs}$ and band 13 in $^{131}\mathrm{La}$ are well reproduced. Moreover, a distinctive rotational mode transition is uncovered in this work, progressing from the principal-axis rotation to the planar rotation and finally to the chiral rotation, as the rotational frequency evolves through different regimes.
%\end{description}
\end{abstract}

\maketitle

%\tableofcontents

\section{\label{sec:1}introduction}
%The investigation of exotic rotational phenomena has been a hot topic in nuclear structure physics for many decades. In the 1950s, Bohr and Mottelson \cite{bohr_mottelson_1953, Bohr1953PhysRev.89.316, Bohr21953PhysRev.90.717.2} introduced the concept that the nucleus undergoes a collective rotation, often referred to as 'electrical rotation', which gives rise to a rotational spectrum characterized by a $\varDelta I  = 2$ transition. An important characteristic of electric rotation bands is the presence of strong $E2$ intraband transitions. Forty years later, 
%In 1997, another entirely new phenomenon, called nuclear chirality,  was proposed . 
%In contrast to conventional rotation observed in well-deformed or superdeformed nuclei, magnetic rotation observed in nearly spherical nuclei and chiral rotation in triaxial nuclei have attracted significant attention in recent decades \cite{HUBEL20051,teng2024systematic,2000article,meng2013progress,sun2019energy}. 
The investigation of rotational phenomena has always been a hot topic in nuclear structure physics for many decades. In the 1950s, Bohr and Mottelson \cite{bohr_mottelson_1953, Bohr1953PhysRev.89.316, Bohr21953PhysRev.90.717.2} introduced the concept of collective nuclear rotation, often referred to as 'electrical rotation', which generates a rotational spectrum with $\varDelta I  = 2$ transitions and is characterized by strong intraband $E2$ transitions. Forty years later, a novel rotational mode distinct from conventional electric rotation, known as magnetic rotation, was proposed~\cite{frauendorf1994proceedings}. 
The mechanism underlying magnetic rotational bands was explained through the shear mechanism \cite{FRAUENDORF1993259}. Several years later, the existence of magnetic rotation was first demonstrated by lifetime measurements for four $M1$ bands in $^{198,199}\mathrm{Pb}$~\cite{clarkPhysRevLett.78.1868}. The experimental characteristics of magnetic rotational bands can be summarized as follows: they exhibit a ${\Delta I} = 1$ rotational structure, featuring significantly enhanced $M1$ transitions and relatively weak $E2$ transitions \cite{clark2000shears, 2001RevModPhys.73.463}. In 1997, another entirely new phenomenon, called nuclear chirality,  was proposed \cite{frauendorf1997tilted}. It consists of a pair of nearly degenerate ${\varDelta} I$ = 1 sequences with the same parity within a nucleus exhibiting triaxial deformation. The existence of nuclear chiral rotation was first experimentally observed in 2001~\cite{starosta2001chiral}. 

To date, approximately 250 candidate magnetic rotational bands have been experimentally reported in the mass regions of $A \approx $ 60, 80, 110, 140, and 190. For more detailed reviews, refer to \cite{meng2013progress,2000article,sun2019energy} and the most recent compilation~\cite{teng2024systematic}. Additionally, over 60 candidate chiral nuclei have been reported in the mass regions of $ A\approx$80, 100, 130, and 190 on the nuclear chart. Detailed data tables can be found in~\cite{xiong2019nuclear}. Notably, the candidate chiral nuclei reported in the $A \approx $ 130 mass region form a significant chiral island. The two exotic rotational modes discussed above, despite the significant differences in their corresponding nuclear deformations, may coexist within a single nucleus. A possible indication of the coexistence of two different rotational modes has been observed in $^{136,137,138}\mathrm{Nd}$ \cite{136Ndarticle,137Ndarticle,138NdPhysRevC.86.044321,1996Pe06PhysRevC.53.R2581,1994De06PhysRevC.49.2990,1997Pe06PETRACHE1997228}. The coexistence of magnetic and chiral rotation within the same nucleus underscores the intricate and interwoven modes of motion. 

In recent investigations, the coexistence of a possible magnetic rotational band and candidate chiral doublet bands has been reported in $^{133}\mathrm{La}$ \cite{petrache2016triaxial}, which is an interesting phenomenon, as its physical mechanism remains to be explored further. Among the isotopes of $\mathrm{La}$, magnetic rotation has been poorly documented. Only certain features suggested in the study of the rotational bands in $^{131}\mathrm{La}$~\cite{1989ef71.PhysRevC.39.471,2000article,131PhysRevLett.58.984,131lAKAUR2017317} align with the characteristics of magnetic rotation. However, the existence of chiral rotation has been claimed in the neighboring $^{128,130,132,133,134}\mathrm{La}$ \cite{128La,130La.Koike:2001ms,132Laarticle,134LaBARK2001577,petrache2016triaxial}. Given that $\mathrm{Cs}$ and $\mathrm{La}$ are $N=74$ isotones, there may be similarities in the properties of their configurations \cite{starosta2001chiral}. To date, evidence for magnetic rotation remains rather limited in $\mathrm{Cs}$ isotopes, with only a few nuclei indicating the potential existence of magnetic rotational bands \cite{131CsKumar2005,131Cs}. In particular, $^{129}\mathrm{Cs}$ was identified as a candidate for magnetic rotation \cite{129Cs2009PhysRevC.79.044317}. In contrast, chiral phenomena have been reported in $^{118,119,121-133}\mathrm{Cs}$ \cite{118Cs,119CsZheng2022,122CsYonNam2005,CsSingh2005,124CsPhysRevC.92.064307,125CsSingh2006,126CsGRODNER201146,126Csli,126Cswang2PhysRevC.75.024309,126CswangPhysRevC.74.017302,129Cs2009PhysRevC.79.044317,130CsSimons_2005,130CsWang,130CsWu_2012,131Cs,132Cs,130La.Koike:2001ms,132cSPhysRevC.68.024318,starosta2001chiral,CsSTAROSTA2001375,chen2023evolution,133CsPhysRevC.110.024301,CsPhysRevC,li2018PhysRevC.97.034306,guo2019PhysRevC.100.034328}. Thus, the investigation into the potential coexistence of magnetic and chiral rotation in $^{129}\mathrm{Cs}$ and $^{131}\mathrm{La}$ presents a particularly fascinating research topic. 

Theoretically, the CDFT has gained much attention by considering Lorentz symmetry in a self-consistent manner and successfully describing a large number of nuclear phenomena in both the ground and the excited state \cite{RING1996193, Vretenar2005RelativisticHT, 119CsZheng2022}. Within the framework of CDFT, the TAC-CDFT has gained widespread application in the study of nuclear magnetic and chiral rotation \cite{meng2013progress}. On the one hand, the TAC-CDFT model based on the point-coupling interaction \cite{1992PhysRevC.46.1757,zhao2010new} has been successfully applied to describe magnetic rotation in $^{60}\mathrm{Ni}$ \cite{zhao2011novel}, $^{58}\mathrm{Fe}$ \cite{steppenbeck2012magnetic}, $^{61}\mathrm{Ni}$ \cite{lin2023PhysRevC.107.014307}, $^{114}\mathrm{In}$ \cite{LI201234}, $^{142}\mathrm{Gd}$ \cite{peng2008covariant}, $^{198,199}\mathrm{Pb}$ \cite{yu2012PhysRevC.85.024318,wang2018PhysRevC.97.064321} and $^{202,203}\mathrm{Bi}$ \cite{2024zhaosuPhysRevC.109.014315}. On the other hand, for chiral rotation, the developed TAC-CDFT model is used to explore the multiple chirality in nuclear rotation for the first time, completely self-consistent and microscopic \cite{zhao2017multiple}. Until now, TAC-CDFT has been used to successfully describe nuclear chiral rotation, such as in $^{106}\mathrm{Ag}$ \cite{zhao2019microscopic}, $^{121-133}\mathrm{Cs}$ \cite{chen2023evolution}, $^{135}\mathrm{Nd}$ \cite{peng2020covariant}, $^{102-107}\mathrm{Rh}$ \cite{2022pengPhysRevC.105.044318}. 

In the present work, the possible rotation coexistence phenomenon in $^{129}\mathrm{Cs}$ and $^{131}\mathrm{La}$ has been investigated by the 3DTAC-CDFT. The evolution of deformation in $^{129}\mathrm{Cs}$ and $^{131}\mathrm{La}$ is investigated. The calculated excitation energies and the relations between the spin and the rotational frequency for the magnetic band in $^{129}\mathrm{Cs}$ and $^{131}\mathrm{La}$ were compared with experimental data. The electromagnetic transition strengths $B(M1)$ and $B(E2)$ values and $B(M1)/B(E2)$ ratios are calculated and compared with the available data. The shear mechanism and the evolution of the orientation angles $\theta$ and $\varphi$ are presented. The angular momentum contributions from protons and neutrons in the ${h}_{11/2}$ and $(gd)$ shells, the possible chiral mechanism has been identified in $^{129}\mathrm{Cs}$ and $^{131}\mathrm{La}$.

\section{\label{sec:2}THEORETICAL FRAMEWORK}
Covariant density functional theory starts from a Lagrangian, and the corresponding Kohn–Sham equations have the form of a Dirac equation with effective fields $S(\mathbf{r})$ and $ {V}^{\mathcal{\mu }} (\mathbf{r})$ derived from this Lagrangian \cite{RING1996193, MENG2006470, NIKSIC2011519, Vretenar2005RelativisticHT,doi:10.1142/9872}. In the 3DTAC-CDFT method, these fields are calculated in the intrinsic frame rotating with a constant angular velocity vector $\boldsymbol{\omega}$ in the Dirac equation as
\begin{equation}
[\boldsymbol{\alpha} \cdot(\boldsymbol{p}-\boldsymbol{V})+\beta(m+S)+V-\boldsymbol{\omega} \cdot \hat{\boldsymbol{J}}] \varPsi_{k}=\epsilon_{k} \varPsi_{k},
\label{eq1}
\end{equation}
where $ \boldsymbol{\hat{J}}$ is the total angular momentum of the nucleon spinors, and the $S$, $V$, and $ \boldsymbol {V} $ are relativistic scalar fields, the time-like component of the vector field, and the space-like components of the vector field, respectively, which are in turn coupled with the nucleon densities and current distributions. The Dirac equation is solved in a set of three-dimensional harmonic oscillator basis iteratively, and one finally obtains the single-nucleon spinors ${\varPsi_k}$, the single-particle Routhians ${\epsilon_k}$, the total energies, the expectation values of the angular momenta, transition probabilities, and so on. The magnitude of the angular velocity $\boldsymbol{\omega}$ is connected to the angular momentum quantum number ${I}$ by the semiclassical relation $\langle\widehat{\boldsymbol{J}}\rangle\cdot\langle\widehat{\boldsymbol{J}}\rangle = I\left (I+1  \right)$. Meanwhile, the orientation of $\boldsymbol{\omega}$ is determined self-consistently by minimizing the total Routhian. 

From the Dirac equation, physical observables such as the quadrupole moment, magnetic moment, and electromagnetic transition probabilities $B(M1)$ and $B(E2)$ can be calculated. The quadrupole moments $Q_{20}$ and $Q_{22}$ are calculated by
\begin{equation}
\begin{aligned}
Q_{20} & =\sqrt{\frac{5}{16 \pi}}\left\langle 3 z^2-r^2\right\rangle = \frac{3 A}{4 \pi} R_0^2 a_{20}, \\
Q_{22} & =\sqrt{\frac{15}{32 \pi}}\left\langle x^2-y^2\right\rangle= \frac{3 A}{4 \pi} R_0^2 a_{22},
\label{eq3}
\end{aligned}
\end{equation}
with $R_{0} = 1.2A^{1/3}$ fm. Among them, when calculating the magnetic rotation, $Q_{22}=0$.  The deformation parameters $\beta$ and $\gamma$ can be obtained
 \begin{equation}
     \beta =\sqrt{a_{20}^{2}+2a_{22}^{2} } 
   ,  \quad   \gamma =\arctan \left [ \sqrt{2} \frac{a_{22} }{a_{20} }  \right ]. 
     \label{eq5}
      \end{equation}
The nuclear magnetic moment is calculated in a relativistic way from the nuclear currents~\cite{Li2009Sci.ChinaSer.G1586, Li2013Phys.Rev.C064307}
\begin{equation}
   \boldsymbol{\mu } =\sum_{k>0} n_{k} \int d^{3}r\left [\frac{mc^{2} }{\hbar c}q\mathit{\Psi_ {k}^{\dagger} }\boldsymbol{r} \times  \boldsymbol{\alpha  }\mathit{ \Psi_ {k}} 
+\kappa\mathit{ \Psi_ {k}^{\dagger}} \beta \boldsymbol{\Sigma} \mathit{\Psi_ {k}}  \right ] 
\end{equation}
where the charge $q$  is 1 for protons and 0 for neutrons in units of $e$. The free anomalous gyromagnetic ratio $\kappa $ is used for the nucleon ($\kappa _{p}$ = 1.739 and $\kappa _{n} = -1.193$).
$\boldsymbol{L}$ and $\boldsymbol{\Sigma}$ are, respectively, the orbital angular momentum and spin \cite{zhao2010new}. In the following calculations, the ratio $mc^{2} /{\hbar c}$ in Eq.~(\ref{eq5}) is taken as 1, as in Refs. \cite{zhao2017multiple,2017wangykPhysRevC.96.054324}. 
The transition probabilities $B(M1)$ and $B(E2)$ are calculated in the semiclassical approximation,
\begin{equation}
\begin{aligned}
B(M 1)= & \frac{3}{8 \pi}\left\{\left[-\mu_z \sin \theta+\cos \theta\left(\mu_x \cos \varphi+\mu_y \sin \varphi\right)\right]^2\right. \\
& \left.+\left(\mu_y \cos \varphi-\mu_x \sin \varphi\right)^2\right\}, \\
B(E 2)= & \frac{3}{8}\left[Q_{20}^p \sin ^2 \theta+\sqrt{\frac{2}{3}} Q_{22}^p\left(1+\cos ^2 \theta\right) \cos 2 \varphi\right]^2 \\
& +\left(Q_{22}^p \cos \theta \sin 2 \varphi\right)^2,
\label{eq6}
\end{aligned}
\end{equation}
where $Q_{20}^{p}$ and $Q_{22}^{p}$ are the quadrupole moments of protons, and $\theta$ and $\varphi$ are the orientation angles of the total angular momentum in the intrinsic frame.

In the present work, the PC-PK1 point coupling density functional \cite{zhao2010new} is adopted. The Dirac equation is solved with a spherical harmonic oscillator basis that has $10$ major shells. The pairing correlation is neglected in the calculations, but one should bear in mind that the pairing correlation \cite{2015zhaoPhysRevC.92.034319,wang2018PhysRevC.97.064321,2017wangykPhysRevC.96.054324,WANG2023137923,2025lvPhysRevC.111.014321,pairingPhysRevC.75.044307} could influence the descriptions of the critical frequency \cite{criticalPhysRevC.73.054308,criticalPhysRevLett.93.052501,WANG2023137923,WUcrit2024138445,ChencriPhysRevC.111.024302} as well as the total angular momentum and ${ B(M1)}$ values. 

\section{\label{sec:3} Results and Discussion}

For the nuclei $^{129}\mathrm{Cs}$ and $^{131}\mathrm{La}$, the possible magnetic bands B8 and 13 have been suggested in Refs. \cite{129Cs2009PhysRevC.79.044317,1989ef71.PhysRevC.39.471} with the same configuration $\pi {h}_{11/2}^{1}\otimes \nu {h}_{11/2}^{-2}$. In the 3DTAC-CDFT calculations, the valence nucleon configurations $\pi {{h}_{11/2}^{1}({g}_{7/2}/{d}_{5/2})^4}\otimes \nu {h}_{11/2}^{10}({g}_{7/2}/{d}_{5/2})^{14}$ and $\pi {{h}_{11/2}^{1}({g}_{7/2}/{d}_{5/2})^6}\otimes \nu {h}_{11/2}^{10}({g}_{7/2}/{d}_{5/2})^{14}$ are obtained, referred to as config1 and config2, respectively. The valence protons of $^{129}\mathrm{Cs}$ and $^{131}\mathrm{La}$ occupy the high-$j$ ${h}_{11/2}$ orbitals, whereas the neutrons populate the high-$j$ ${h}_{11/2}$ orbitals or ${s}_{1/2}/{d}_{3/2}$ orbitals. Therefore, it is intriguing to discuss the rotational mechanisms that may arise from the neutron occupancy. As mentioned in Ref. \cite{zhao2017multiple}, for $^{129}\mathrm{Cs}$ and $^{131}\mathrm{La}$, Eq.~(\ref{eq1}) was solved iteratively by self-consistently distributing nucleons into single-particle orbitals based on their energy levels, starting from the lowest energy state. The self-consistent calculation produces a nucleon configuration where one proton occupies the bottom of the ${h}_{11/2}$ shell, eight neutrons populate the top of the ${h}_{11/2}$ shell, with two in antialigned coupling, and two remaining neutrons fill the ${s}_{1/2}/{d}_{3/2}$ orbitals above the $Z$ = 50 shell closure. In short, the calculated configuration of the valence nucleons $\pi {{h}_{11/2}^{1}({g}_{7/2}/{d}_{5/2})^4}\otimes \nu {h}_{11/2}^{8}({g}_{7/2}/{d}_{5/2})^{14}({s}_{1/2}/{d}_{3/2})^{2}$ in $^{129}\mathrm{Cs}$ is labeled as config3 and $\pi {{h}_{11/2}^{1}({g}_{7/2}/{d}_{5/2})^6}\otimes \nu {h}_{11/2}^{8}({g}_{7/2}/{d}_{5/2})^{14}({s}_{1/2}/{d}_{3/2})^{2}$ in $^{131}\mathrm{La}$ is labeled as config4. They have identical unpaired nucleon configuration, i.e., $\pi {h}_{11/2}^{1}\otimes \nu {h}_{11/2}^{-2}$.

\begin{figure}
\centering
\includegraphics[width=8cm]{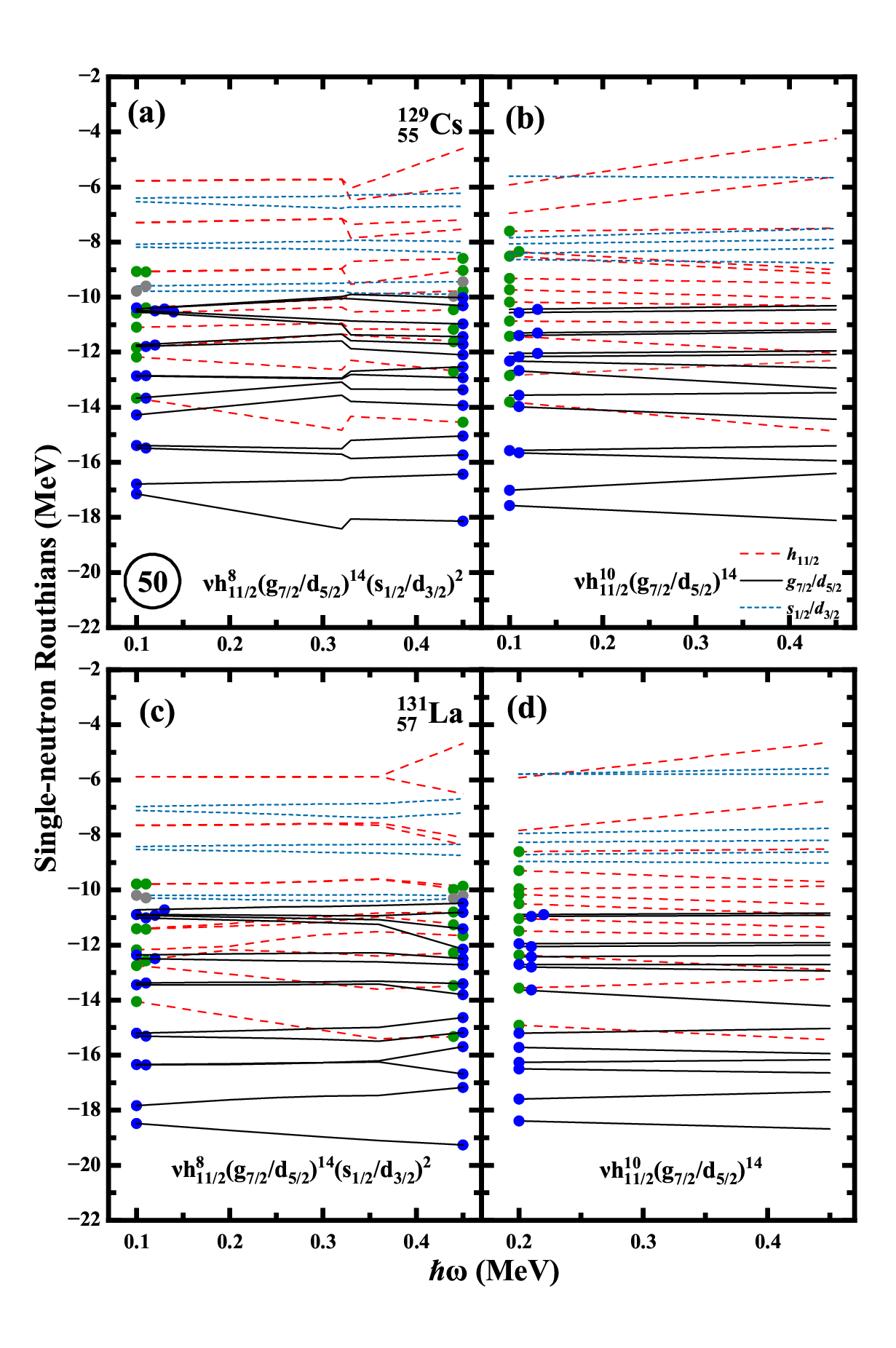}% Here is how to import EPS art
\caption{\label{fig:wide}Single-neutron Routhians near the Fermi surface for $^{129}\mathrm{Cs}$ and $^{131}\mathrm{La}$ as a function of the rotational frequency $\hbar\omega$ for configuration $\pi {h}_{11/2}^{1}\otimes \nu {h}_{11/2}^{-2}$. The black solid lines and blue dotted lines represent the positive-parity levels of the $\nu {g}_{7/2}/{d}_{5/2}$ and $\nu {s}_{1/2}/{d}_{3/2}$ orbitals, respectively, while the red dashed lines denote the negative-parity levels of the $\nu{h}_{11/2}$ orbitals. The occupied single-neutron states are denoted by colored solid dots: green for the $\nu{h}_{11/2}$ orbitals, blue for the $\nu {g}_{7/2}/{d}_{5/2}$ orbitals, and gray for the $\nu {s}_{1/2}/{d}_{3/2}$ orbitals. }\label{fig1}
\label{fig1(a)-(d)}

\end{figure}

% On the one hand, regarding neutron occupancy in isotopes $^{129}\mathrm{Cs}$ and $^{131}\mathrm{La}$, the nuclear structure exhibits distinct characteristics depending on the filling of the ${h}_{11/2}$ orbitals. With eight neutrons, corresponding to a near half-filled shell, strong collective nucleon motion emerges, facilitating the formation of stable triaxial deformation. In contrast, with ten neutrons approaching a fully-filled shell, the system transitions to predominantly single-particle behavior, reducing collectivity and favoring axially symmetric nuclear shapes. This latter configuration is particularly conducive to the development of magnetic rotation. On the other hand, regarding angular momentum contribution, chiral rotation requires triaxial nuclear deformation, with the total angular momentum arising from rotations about the $s$, $m$ and $l$ axes. For eight neutrons in the ${h}_{11/2}$ orbitals, the nuclear angular momentum distribution becomes significantly more uniform, with substantial contributions from all three principal axes, thereby enabling chiral rotation. In contrast, magnetic rotation primarily originates from single-particle angular momentum contributions in the high-$j$ ${h}_{11/2}$ orbitals. With ten neutrons occupying the ${h}_{11/2}$ orbitals, the single-particle contribution dominates, while collective rotational contributions become comparatively weaker.  

\subsection{Shape coexistence } 
Figure \ref{fig1} displays the single-neutron Routhians for config1, config2, config3, and config4 as a function of rotational frequency $\hbar\omega$. The different occupation patterns of neutrons lead to varying degrees of hole-like character, thus giving rise to distinct rotational modes. As shown in Fig. \ref{fig1}, there are twenty-four neutrons above the $N=50$ shell closure. The ${g}_{7/2}/{d}_{5/2}$ orbitals are fully occupied, with the remaining nucleons distributed between the ${h}_{11/2}$ orbitals and the ${s}_{1/2}/{d}_{3/2}$ orbitals, respectively. In Fig. \ref{fig1}(a) and \ref{fig1}(c), eight neutrons occupy the ${h}_{11/2}$ orbitals, while the remaining two neutrons populate the ${s}_{1/2}/{d}_{3/2}$ orbitals. In contrast, Fig. \ref{fig1}(b) and \ref{fig1}(d) demonstrate all remaining ten neutrons occupy the ${h}_{11/2}$ orbitals. The energy amplitude range from 0.1 MeV or 0.2 MeV to 0.45 MeV is presented in Fig. \ref{fig1}. However, sometimes when the rotation frequency exceeds a certain range, the solution will diverge. Figure \ref{fig1}(a) and \ref{fig1}(c) show the occupation of single-neutron Routhians at different rotational frequencies. A sudden change in the single-particle levels occurs when the rotational frequency reaches a critical value  \cite{criticalPhysRevC.73.054308,criticalPhysRevLett.93.052501,WANG2023137923,WUcrit2024138445,ChencriPhysRevC.111.024302}, which can be attributed to the transition from planar rotation to nonplanar rotation. A detailed discussion is presented in Sec. C.

\begin{figure}
\centering
\includegraphics[width=8cm]{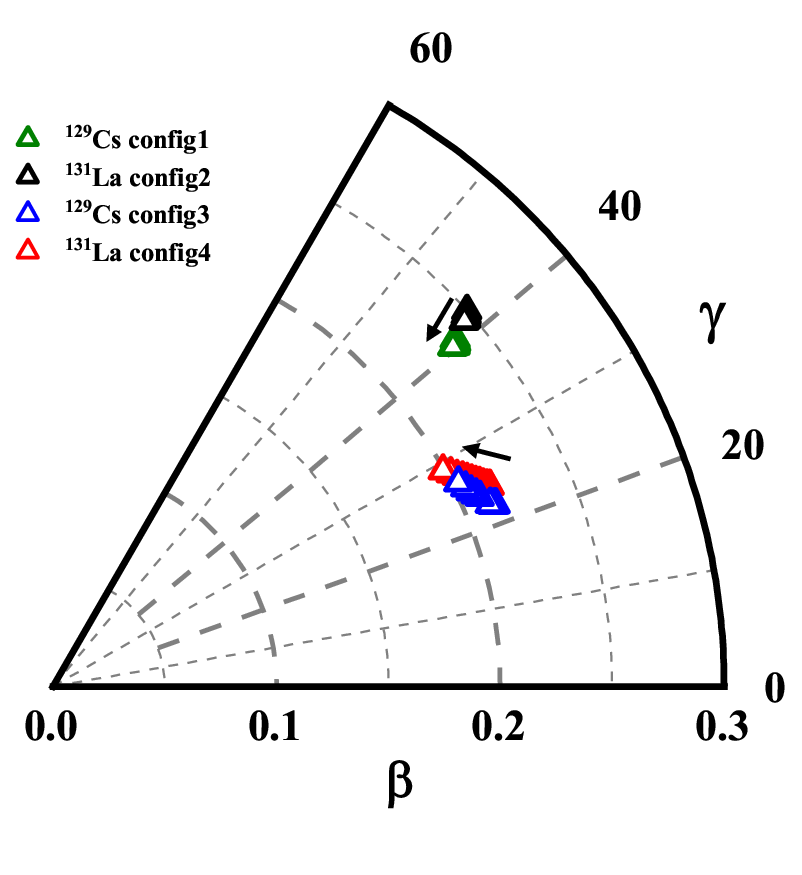}% Here is how to import EPS art
\caption{\label{fig:wide}The evolutions of deformation in the ($\beta$, $\gamma$) plane driven by rotation in the 3DTAC-CDFT calculations for config1, config2, congfig3, and config4, respectively. The arrows indicate the increasing direction of the rotational frequency $\hbar\omega$.}
\label{fig2}
\end{figure}

Within the $A\approx130$ mass region, nuclei lie in the transitional area between nearly spherical and deformed nuclei \cite{Guo2020Collective}. There may also be competition among various alignments in some nuclei, exhibiting rich nuclear structure phenomenas, such as shape coexistence \cite{shape1AYANGEAKAA2016254,shape2PETRACHE2019241,shape3104AgPhysRevC.88.024306,shape4LEONI2024104119,shape5atoms11090117}. Figure \ref{fig2} depicts the evolution of the deformation parameters $\beta$ and $\gamma$ in $^{129}\mathrm{Cs}$ and $^{131}\mathrm{La}$ as a function of increasing rotational frequency $\hbar\omega$ in the 3DTAC-CDFT calculations. For the oblate configurations (config1 and config2), the deformation parameters ($\beta$, $\gamma$) are (0.23, 41$^\circ$) for band B8 in $^{129}\mathrm{Cs}$ and (0.25, 42$^\circ$) for band 13 in $^{131}\mathrm{La}$. With an increasing rotational frequency for config1 in $^{129}\mathrm{Cs}$ and config2 in $^{131}\mathrm{La}$, the $\beta$ values, which typically range between 0.20 and 0.30, show a gradual decrease. Meanwhile, the $\gamma$ values remain stable at approximately 40$^\circ$. In the case of possible chiral rotation, the deformation parameters  $\beta$ are around 0.20 for config3 and config4. Simultaneously, with increasing rotational frequency, the $\gamma$ deformation for both config3 and config4 evolves toward $\ gamma\approx30^\circ$ at higher frequencies. It is worth noting that these four configurations may lead to the formation of distinct rotational modes. 

\subsection{Magnetic rotation}

\begin{figure}
\centering
\includegraphics[width=6cm]{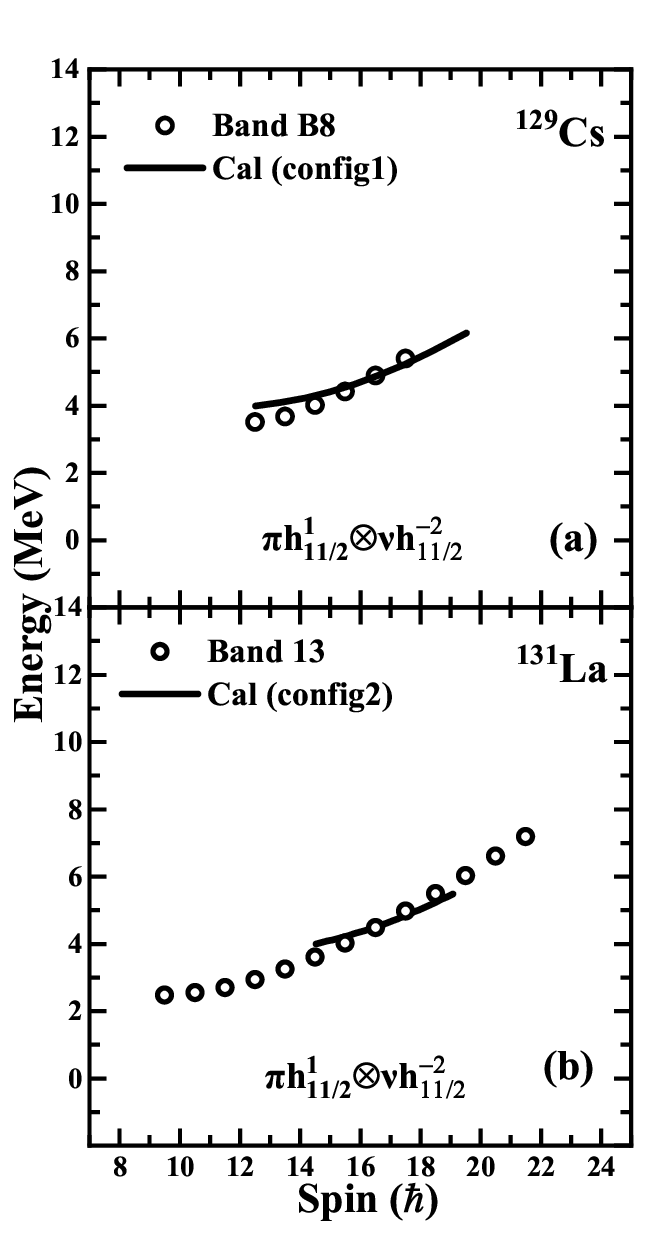}% Here is how to import EPS art
\caption{\label{fig:wide}The calculated energy spectra as a function of spin by 3DTAC-CDFT calculations in comparison with the experimental data for (a) band B8  in $^{129}\mathrm{Cs}$ and (b) band 13 in $^{131}\mathrm{La}$. The corresponding experimental data are taken from Refs.  \cite{129Cs2009PhysRevC.79.044317,1989ef71.PhysRevC.39.471}.
}\label{fig3}
\end{figure}

\begin{figure}
\centering
\includegraphics[width=6cm]{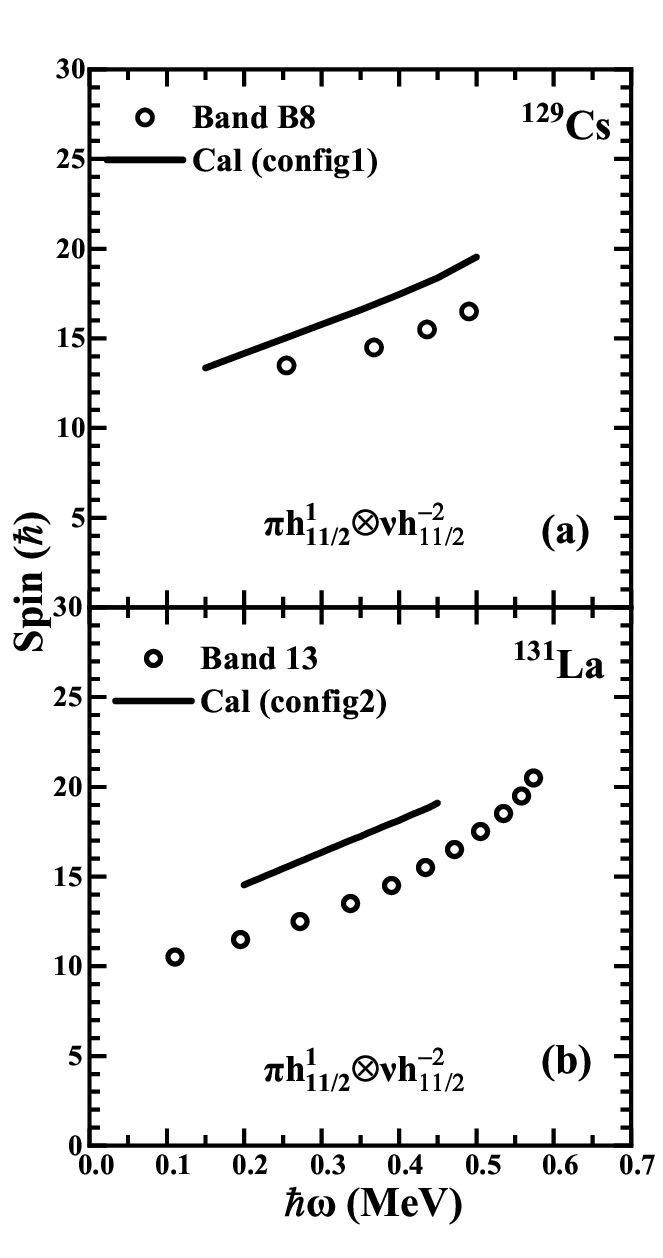}% Here is how to import EPS art
\caption{\label{fig:wide}The spin as a function of the rotational frequency $\hbar\omega$ in the 3DTAC-CDFT calculations compared with the experimental data \cite{129Cs2009PhysRevC.79.044317,1989ef71.PhysRevC.39.471} for (a) band B8 in$^{129}\mathrm{Cs}$  and (b) band 13 in $^{131}\mathrm{La}$.}\label{fig4}
\end{figure}

Figure \ref{fig3} shows the comparison between the experimental \cite{129Cs2009PhysRevC.79.044317,1989ef71.PhysRevC.39.471} and calculated energies for band B8 in $^{129}\mathrm{Cs}$ and band 13 in $^{131}\mathrm{La}$ with configuration $\pi {h}_{11/2}^{1}\otimes \nu {h}_{11/2}^{-2}$. The calculated excitation energies show excellent agreement with the experimental data \cite{129Cs2009PhysRevC.79.044317,1989ef71.PhysRevC.39.471} as a function of spin. As illustrated in Fig. \ref{fig4}, the calculated spin values show overestimation compared to experimental data \cite{129Cs2009PhysRevC.79.044317,1989ef71.PhysRevC.39.471}. Despite this, the overall trends remain consistent, with this deviation most likely originating from pairing correlation effects \cite{2015zhaoPhysRevC.92.034319,wang2018PhysRevC.97.064321,2017wangykPhysRevC.96.054324,WANG2023137923,2025lvPhysRevC.111.014321,pairingPhysRevC.75.044307}. 

\begin{figure*}
\centering
\includegraphics[width=12cm]{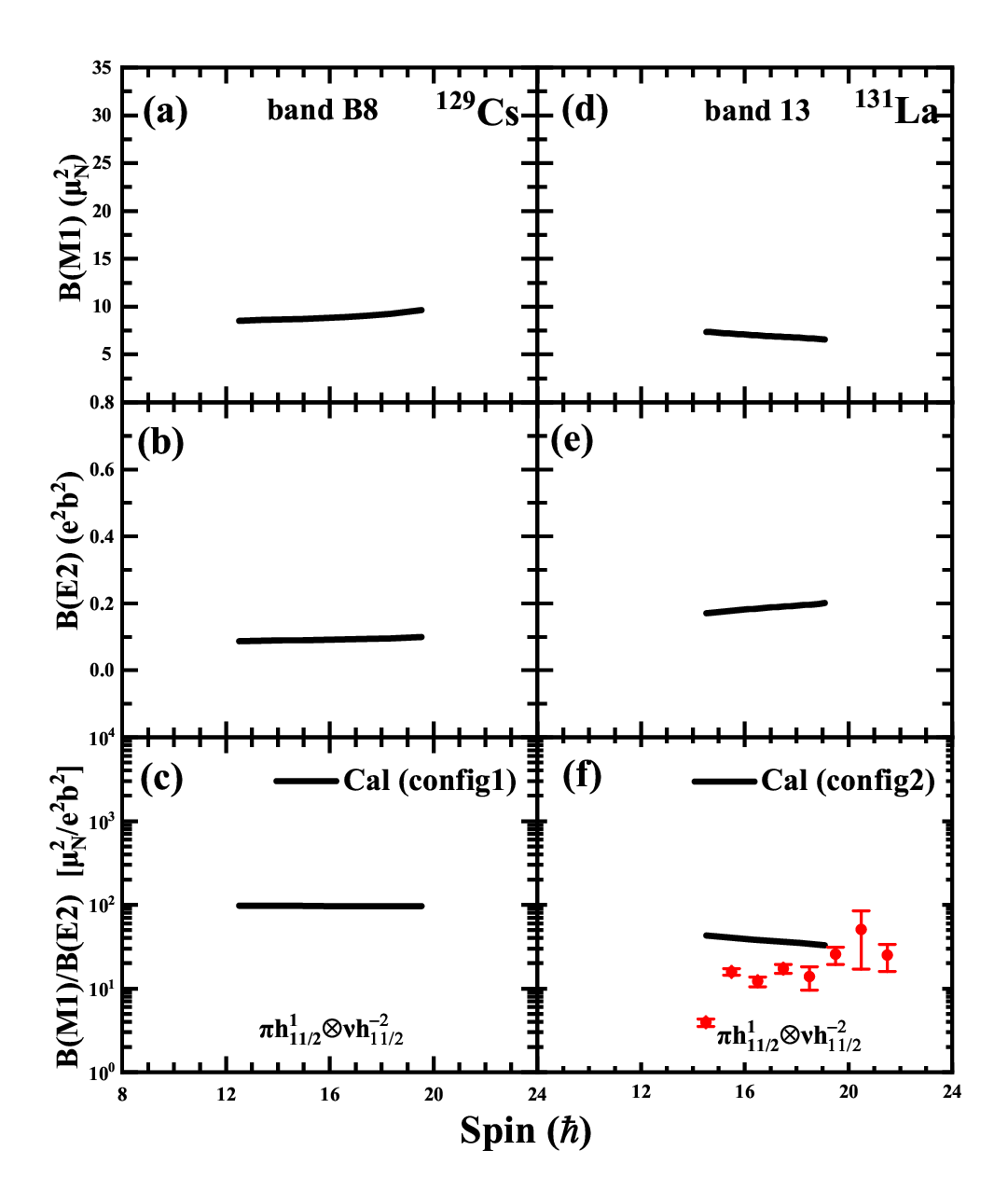}% Here is how to import EPS art
\caption{\label{fig:wide} The transition probabilities $B(M1)$ and $B(E2)$ values, and $B(M1)/B(E2)$ ratios as a function of spin for the band B8 in $^{129}\mathrm{Cs}$  and band 13 in $^{131}\mathrm{La}$ by the 3DTAC-CDFT calculations. The corresponding experimental data are taken from Refs. \cite{1989ef71.PhysRevC.39.471}.}
\label{fig5}
\end{figure*}

\begin{figure*}
\centering
\includegraphics[width=14cm]{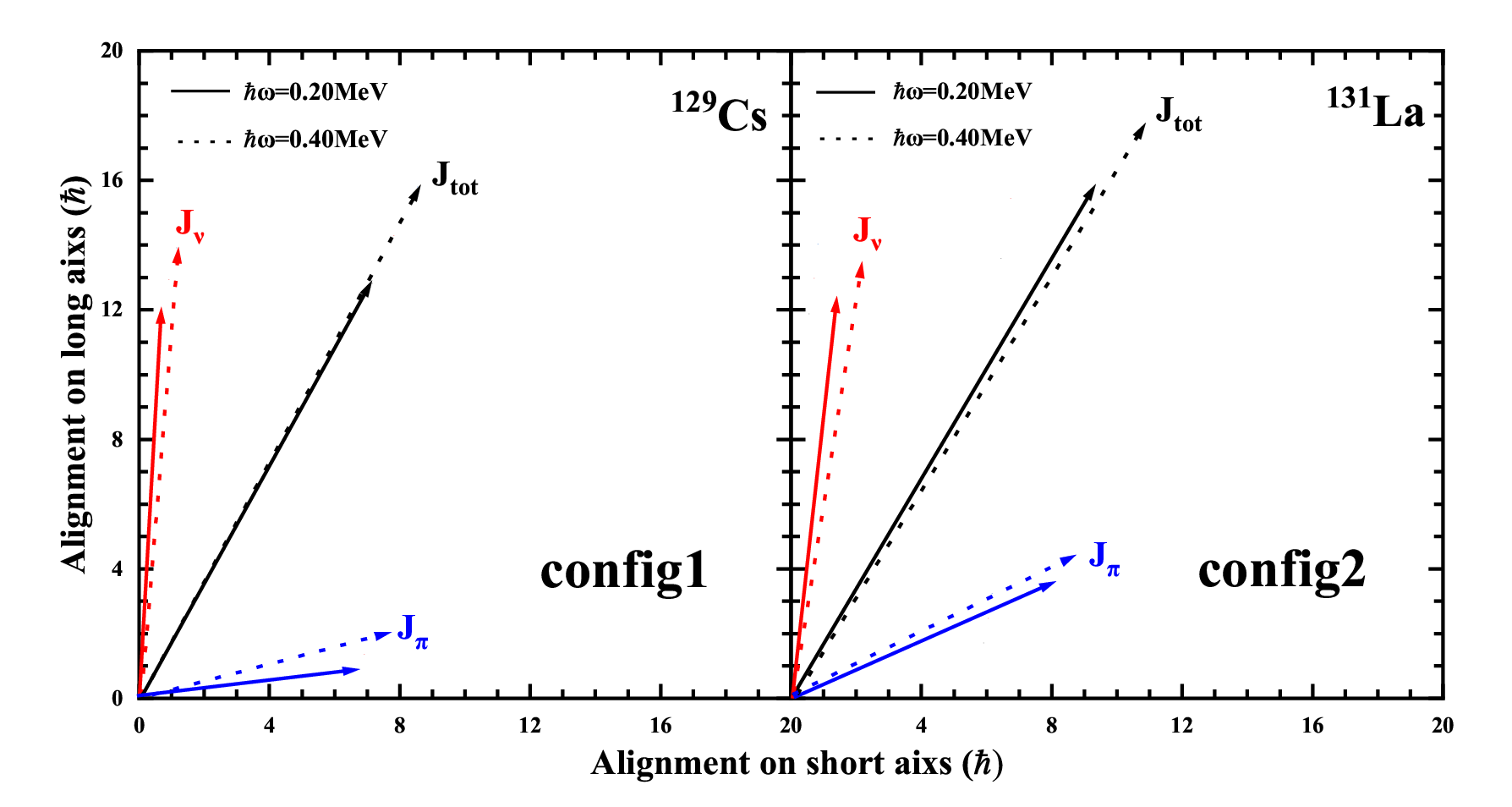}% Here is how to import EPS art
\caption{\label{fig:wide}Composition of the total angular momentum at two
 different rotational frequencies in 3DTAC-CDFT calculation for $^{129}\mathrm{Cs}$ with config1 and $^{131}\mathrm{La}$ with config2.}\label{fig6}
\end{figure*}

The characteristic features of magnetic rotation bands include strongly enhanced $M1$ transitions at low spins, a gradual decrease in $M1$ with increasing spin, and relatively weak $E2$ transitions compared to the dominant $M1$ transitions. Figure \ref{fig5} presents the calculated $B(M1)$ and $B(E2)$ values and $B(M1)/B(E2)$ ratios as a function of the total angular momentum for $\pi {h}_{11/2}^{1}\otimes \nu {h}_{11/2}^{-2}$. The calculated results are compared with experimental data from Ref. \cite{1989ef71.PhysRevC.39.471}. For config2 in $^{131}\mathrm{La}$, the 3DTAC-CDFT calculations demonstrate a clear decreasing trend in $B(M1)$ values with increasing spin, as shown in Fig. \ref{fig5}. In contrast, config1 in $^{129}\mathrm{Cs}$ exhibits stable $B(M1)$ values without significant reduction, attributed to minimal variations in the shear angle ${\Theta}$ (see Fig. \ref{fig6}). In contrast to the large $B(M1)$ values, the $B(E2)$ values for both bands obtained from 3DTAC-CDFT calculations remain consistently small, showing nearly constant or even marginally increasing behavior. As illustrated in Fig. \ref{fig5}, the calculated $B(M1)/B(E2)$ ratios are compared with the values derived from the experimental branching ratios \cite{1989ef71.PhysRevC.39.471}. It can be seen that the calculated results exhibit a slight overestimation compared to the experimental data. Given that no adjustable parameters are introduced in the present calculation, the level of agreement between the theoretical predictions and the experimental observations is considered acceptable \cite{2025lvPhysRevC.111.014321}. Moreover, as shown in Refs. \cite{2017wangykPhysRevC.96.054324,WANG2023137923} the inclusion of pairing correlations tends to reduce the B(M1) values. Therefore, it is expected that the agreement between the calculated results and the data might be further improved after the inclusion of pairing correlations.

Another typical characteristic of magnetic rotation is the shear mechanism \cite{FRAUENDORF1993259}. To examine the shear mechanism in band B8 of $^{129}\mathrm{Cs}$ and band 13 of $^{131}\mathrm{La}$, the evolution of this mechanism at two distinct rotational frequencies is shown in Fig. \ref{fig6}. The angular momentum vectors for proton $\bm{J}_{\pi}$ and neutron $\bm{J}_{\nu}$, as well as the total angular momentum vectors $\bm{J}_{tot} = J_{\pi}+J_{\nu}$ for config1 and config2 are illustrated in Fig. \ref{fig6}. In the magnetic rotational bands of $^{129}\mathrm{Cs}$ and $^{131}\mathrm{La}$, neutrons occupy the ${h}_{11/2}$ orbitals, while protons occupy the ${h}_{11/2}$ and ${d}_{5/2}/{g}_{7/2}$ orbitals. 
For these two bands, the proton particles in ${h}_{11/2}$ preferentially orient their angular momentum along the short axis, while the neutron hole(s) in the ${h}_{11/2}$ shell predominantly align their angular momentum with the long axis. The angular momenta are predominantly contributed by these orbitals. As depicted in Fig. \ref{fig6}, with increasing rotational frequency, the progressive alignment of $\bm{J}_{\pi}$ toward $\bm{J}_{\nu}$ generates a higher angular momentum while maintaining the direction of the total angular momentum vector. The process depicted in Fig. \ref{fig6} can be regarded as the closing motion of a shear pair. These characteristic features indicate that the band B8 in $^{129}\mathrm{Cs}$ and the band 13 in $^{131}\mathrm{La}$ exhibit the properties of magnetic rotational bands.

\begin{figure*}
\centering
\includegraphics[width=14cm]{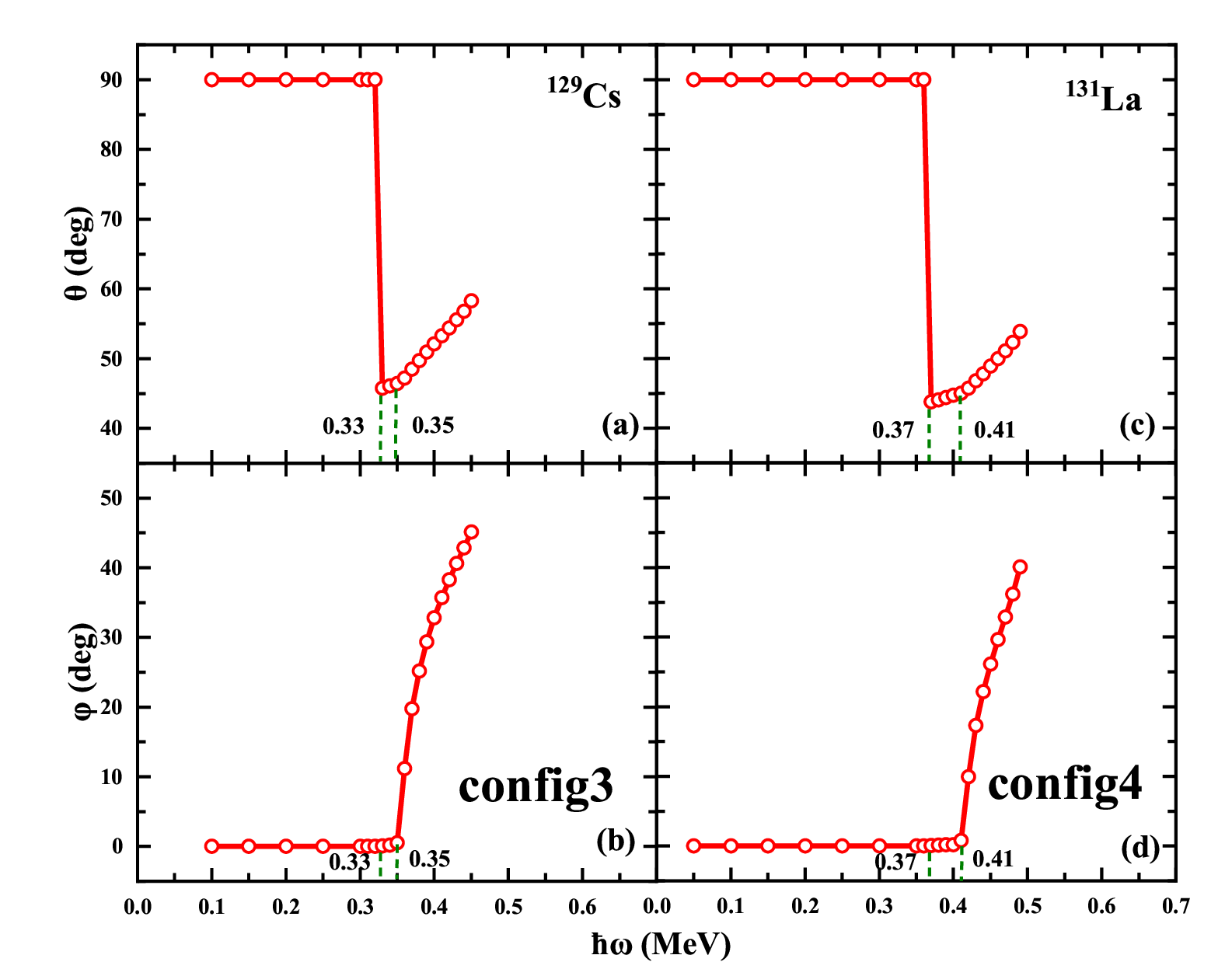}% Here is how to import EPS art
\caption{\label{fig:wide}Evolution of the orientation angles $\theta$ and $\varphi$ for the angular momentum $J$ as driven by the increasing cranking frequency $\omega$ from the  configuration $\pi {h}_{11/2}$ to the configuration  $\pi {h}_{11/2}^{1}\otimes \nu {h}_{11/2}^{-2}$ in $^{129}\mathrm{Cs}$ and $^{131}\mathrm{La}$.}\label{fig7}
\end{figure*}
\subsection{Chiral rotation}
%There are 57 protons and 72 neutrons in the nucleus $^{129}\mathrm{Cs}$ and 57 protons and 74 neutrons in the nucleus in $^{131}\mathrm{La}$.

\begin{figure*}
\centering
\includegraphics[width=14cm]{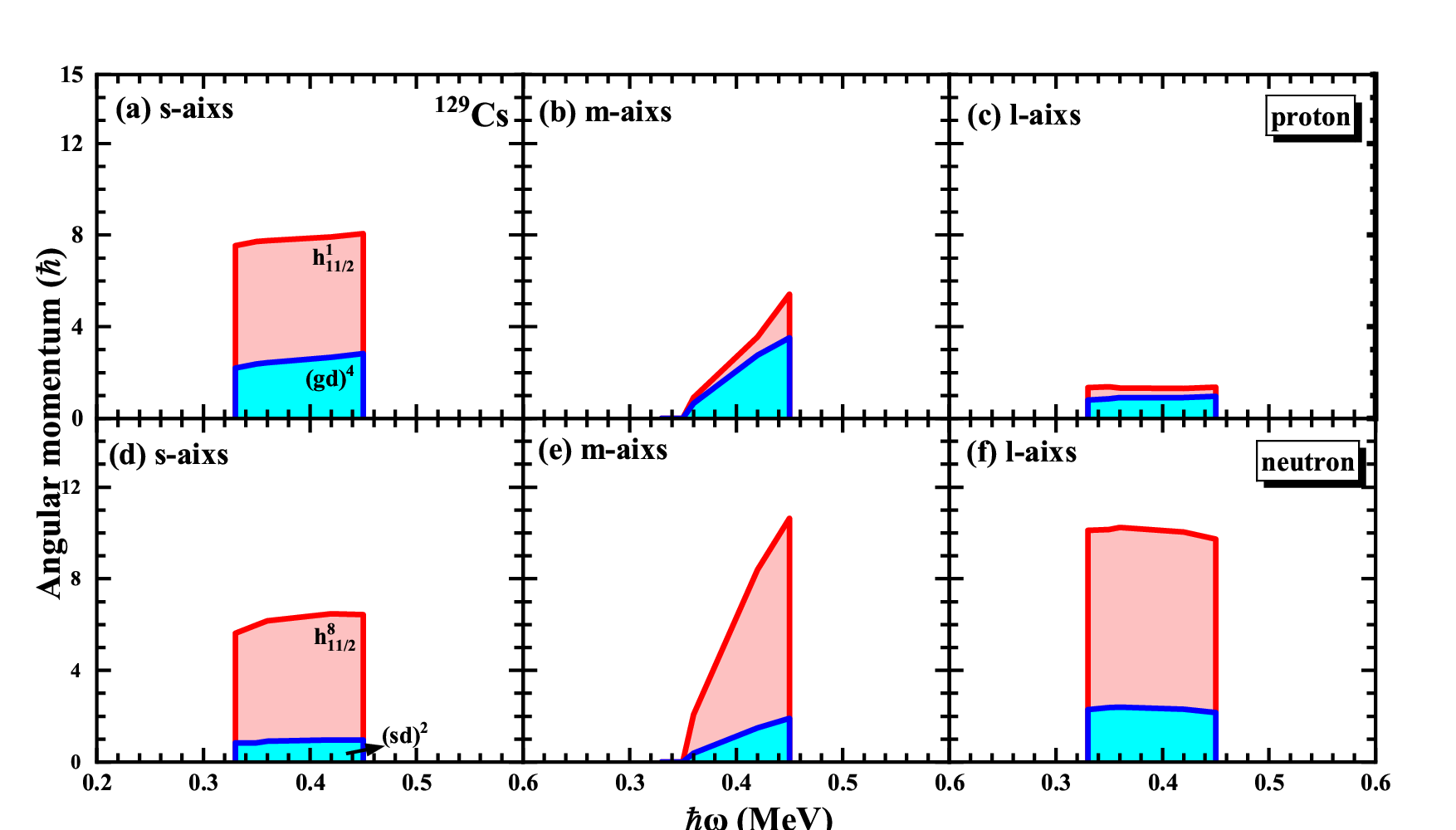}% Here is how to import EPS art
\caption{\label{fig:wide}Contributions of protons in the ${h}_{11/2}$ and ${(gd)}$ orbitals as well as neutrons in the ${h}_{11/2}$ and ${(sd)}$ orbitals to the angular momenta along the short, medium and long axes for the configuration $\pi {h}_{11/2}^{1}\otimes \nu {h}_{11/2}^{-2}$ in $^{129}\mathrm{Cs}$ calculated by 3DTAC-CDFT.}\label{fig8}
\end{figure*}

\begin{figure*}
\centering
\includegraphics[width=14cm]{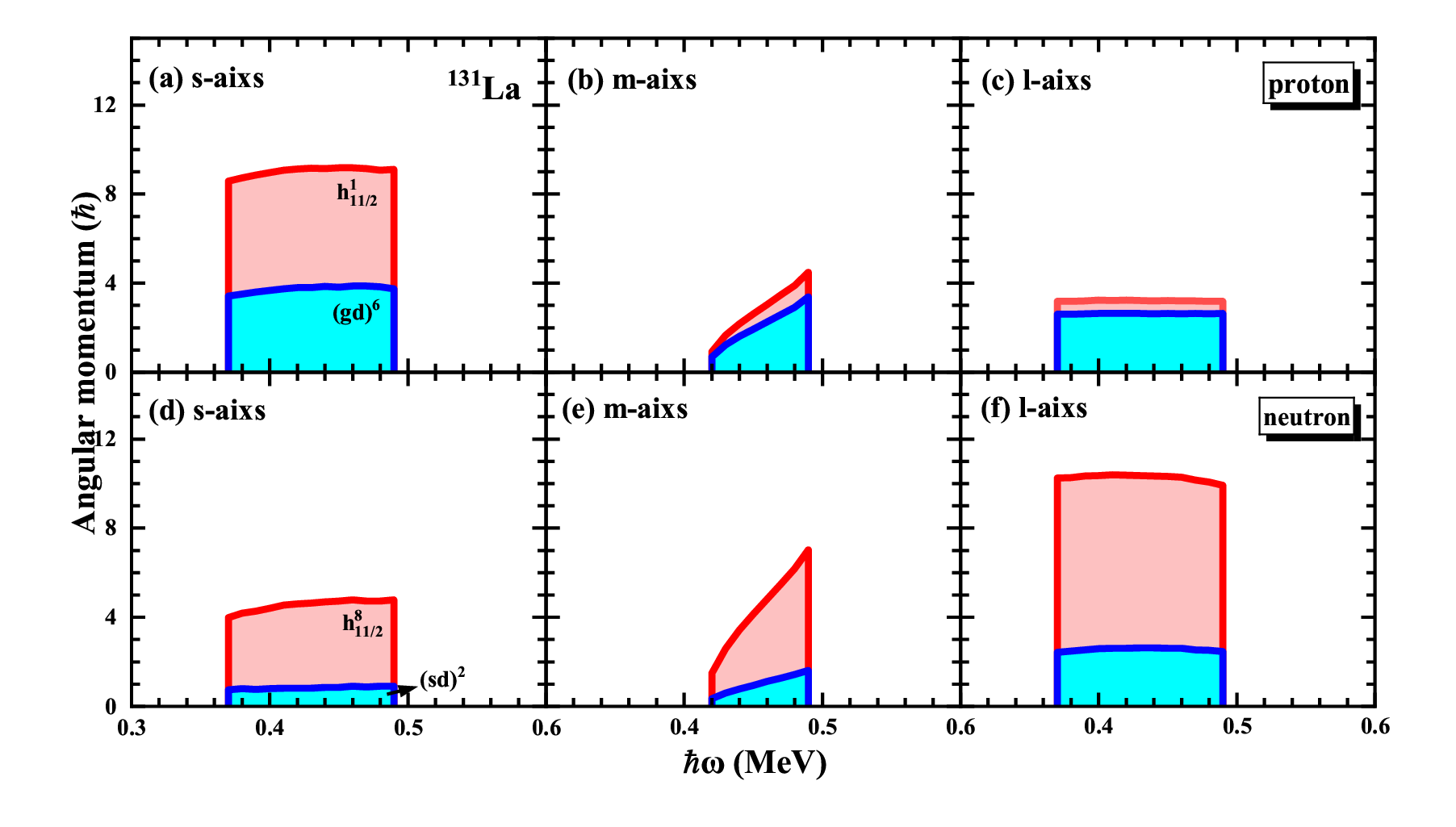}% Here is how to import EPS art
\caption{\label{fig:wide}Contributions of protons in the ${h}_{11/2}$ and ${(gd)}$ orbitals as well as neutrons in the ${h}_{11/2}$ and ${(sd)}$ orbitals to the angular momenta along the short, medium and long axes for the configuration $\pi {h}_{11/2}^{1}\otimes \nu {h}_{11/2}^{-2}$ in $^{131}\mathrm{La}$ calculated by 3DTAC-CDFT.}\label{fig9}
\end{figure*}

To investigate the potential existence of chiral geometry in config3 of $^{129}\mathrm{Cs}$ and config4 of $^{131}\mathrm{La}$, the evolution of orientation angles $\theta$ and $\varphi$ as a function of increasing rotational frequency is illustrated in Fig. \ref{fig7}. It can be observed that in the low-frequency regime, the azimuthal angle $\varphi$ approaches zero and the polar angle $\theta$ maintains 90$^\circ$, indicating a preference for principal-axis rotation at low spins. When the rotational frequency reaches $\hbar\omega $= 0.33 MeV in Fig. \ref{fig7}(a) and $\hbar\omega $= 0.37 MeV in Fig. \ref{fig7}(c), also known as the critical frequency ${\omega}_{crit}$ \cite{criticalPhysRevC.73.054308,criticalPhysRevLett.93.052501,WANG2023137923,WUcrit2024138445,ChencriPhysRevC.111.024302}, the abrupt changes in $\theta$ (from 90° to a finite value) signifies the transition of the system from planar rotation to aplanar rotation \cite{chenPhysRevC.87.024314}. For $\hbar\omega $ ranging from 0.33 to 0.35 MeV in Figs. \ref{fig7}(a) and \ref{fig7}(c), and from 0.37 to 0.41 MeV in Figs. \ref{fig7}(b) and \ref{fig7}(d), this might correspond to the transition of the system from chiral vibration to static chirality \cite{staticPhysRevLett.99.172501}. Beyond $\hbar\omega $ = 0.35 MeV for $^{129}\mathrm{Cs}$ and $\hbar\omega $ = 0.41 MeV for $^{131}\mathrm{La}$, nonzero values of $\varphi$ emerge, indicating the onset of static chirality. This is also related to the sudden appearance of the angular momentum along the $m$ axis at this frequency, as shown in Fig. \ref{fig8} and Fig. \ref{fig9}. The change from planar rotation to aplanar rotation can be observed in the single-particle Routhian diagram. At specific frequencies, some energy levels will split, leading to a change in the position of the particles. This transition from planar to chiral rotation has been clearly explained in both nuclei. Meanwhile, positive-parity doublet bands with three-quasiparticle configurations in odd-A nuclei $^{125-131}\mathrm{Cs}$ were systematically studied in Ref. \cite{guo2019PhysRevC.100.034328}. The configuration $\pi {h}_{11/2}^{1}\otimes \nu {h}_{11/2}^{-1}({s}_{1/2}/{d}_{3/2})^{1}$ was suggested to $^{129}\mathrm{Cs}$. This suggests the possible existence of $M\chi D$ \cite{mxdPhysRevC.73.037303} in $^{129}\mathrm{Cs}$.

%It is planar rotation up to the $\hbar\omega $ = 0.35 MeV, and significant changes occur in $\theta$ and $\varphi$ after this critical frequency $\omega_{crit}$
%The mutations in the values of $\theta$ mark a change in configuration from config2 to config3. It should be noted that the process in which the value of $\varphi$ changes from zero to non-zero value can be used to describe the transformation from dynamic chirality to static chirality of the system. 


The angular momentum contributions along the $s$, $m$ and $l$ axes from protons in the $h_{11/2}$ and ($gd$) orbitals, as well as neutrons in the $h_{11/2}$ and $(sd)$ orbitals for config3 in $^{129}\mathrm{Cs}$ and config4 in $^{131}\mathrm{La}$  are presented in Fig. \ref{fig8} and Fig. \ref{fig9}. As illustrated, the contribution of the proton to the angular momentum originates primarily from a single proton particle at the bottom of the $h_{11/2}$ shell, which contributes approximately 5.5$\hbar$ along the $s$ axis. Gleichzeitig, the core ($gd$) orbitals contribute a partial component of the angular momentum through collective rotation. In contrast, neutrons (including core contributions) in the $h_{11/2}$ orbitals contribute approximately 8$\hbar$ along the $l$ axis. Collective effects also influence these neutrons, resulting in additional contributions along the $s$ and $m$ axes. The presence of angular momentum along the $m$ axis gives rise to a critical frequency that emerges. These observations suggest potential candidates for chiral rotation in both $^{129}\mathrm{Cs}$ and $^{131}\mathrm{La}$.

\section{\label{sec:4}SUMMARY AND PROSPECTS}

%The structure of odd-$A$ nuclei in the $A \approx130$ mass region exhibits remarkable complexity. This intricate nature renders the investigation of these nuclei both theoretically challenging and crucial for advancing nuclear structure theory. 
3DTAC-CDFT calculations are used to investigate the coexistence of possible magnetic and chiral rotational modes in $^{129}\mathrm{Cs}$ and $^{131}\mathrm{La}$. The single-neutron Routhians near the Fermi surface in $^{129}\mathrm{Cs}$ and $^{131}\mathrm{La}$ reveal the microscopic mechanism underlying neutron occupation, offering enhanced understanding of the formation processes for both possible magnetic and chiral rotation. To elucidate the nature of the magnetic dipole band based on the configuration $\pi {h}_{11/2}^{1}\otimes \nu {h}_{11/2}^{-2}$ in $^{129}\mathrm{Cs}$ and $^{131}\mathrm{La}$, experimental excitation energies, the relation between spin and rotational frequency, deformation parameters, $B(M1)$ and $B(E2)$ values and $B(M1)/B(E2)$ ratios were compared with 3DTAC-CDFT calculations. The theoretical results demonstrate good agreement with experimental observations. In addition, the shear mechanism is explicitly revealed in calculations through frequency-dependent decomposition of total angular momenta into proton and neutron contributions at different frequencies. This study provides strong evidence for identifying  band B8 in $^{129}\mathrm{Cs}$ \cite{129Cs2009PhysRevC.79.044317} and band 13 in $^{131}\mathrm{La}$ \cite{1989ef71.PhysRevC.39.471} as magnetic rotation candidates.

The possible chiral rotation in $^{129}\mathrm{Cs}$ and $^{131}\mathrm{La}$ can be understood through the evolution of $\theta$ and $\varphi$ angles and the corresponding angular momentum contributions. The transition of $\varphi$ from zero to non-zero values signifies the transition from planar rotation to aplanar rotation, while the emergence of the critical frequency indicates the evolution from dynamic to static chirality. Moreover, current theoretical predictions suggest the possible existence of  $M\chi D$ in $^{129}\mathrm{Cs}$, while experimental verification of these predicted chiral bands requires further investigation. Theoretical calculations in this work demonstrate the possible coexistence of magnetic rotation and chiral rotation in nuclei within the $A\approx130$ mass region, and elucidate the underlying microscopic mechanisms governing their formation. Nevertheless, a deeper understanding of the band structures and a thorough analysis of rotational modes in the $ A\approx130$ mass region remain essential for further progress. 

\begin{acknowledgments}
The authors thank Prof. Q. B. Chen and Y. K. Wang for helpful discussions. This research was funded by the National Natural Science Foundation of China (Nos. 12475119 and 11675063) and the Scientific Research Project of the Education Department of Jilin Province (No. JJKH20241242KJ). 
%Natural Science Foundation of Jilin Province (No.20220101017JC), and Key Laboratory of Nuclear Data Foundation (JCKY2020201C157).
\end{acknowledgments}

\nocite{*}

\bibliography{apssamp}% Produces the bibliography via BibTeX.

\end{document}